# Cross-lingual Data Transformation and Combination for Text Classification

*Jun Jiang, Shumao Pang, Xia Zhao, Liwei Wang, Andrew Wen, Hongfang Liu\*, Qianjin Feng\**

*Abstract*—Text classification is a fundamental task for text data mining. In order to train a generalizable model, a large volume of text must be collected. To address data insufficiency, cross-lingual data may occasionally be necessary. Cross-lingual data sources may however suffer from data incompatibility, as text written in different languages can hold distinct word sequences and semantic patterns. Machine translation and word embedding alignment provide an effective way to transform and combine data for cross-lingual data training. To the best of our knowledge, there has been little work done on evaluating how the methodology used to conduct semantic space transformation and data combination affects the performance of classification models trained from cross-lingual resources. In this paper, we systematically evaluated the performance of two commonly used CNN (Convolutional Neural Network) and RNN (Recurrent Neural Network) text classifiers with differing data transformation and combination strategies. Monolingual models were trained from English and French alongside their translated and aligned embeddings. Our results suggested that semantic space transformation may conditionally promote the performance of monolingual models. Bilingual models were trained from a combination of both English and French. Our results indicate that a cross-lingual classification model can significantly benefit from cross-lingual data by learning from translated or aligned embedding spaces.

## I. INTRODUCTION

TEXT classification, which classifies documents into a taxonomy of categories, has been widely used for various text mining applications[1-4]. Feature representations and a labeled training dataset are required to train a deliverable model. With the advancement of deep learning, word embeddings, in which words are represented as context related semantic vectors, have become a popular feature representation approach for text classification tasks[5-7]. It can, however, be labor intensive to obtain enough training data for training.

One potential solution is to leverage cross-lingual labeled datasets (i.e., labeled datasets from other languages) but may suffer from data incompatibility as text written in different languages can hold distinct word sequences and semantic patterns. In order to leverage cross-lingual resources for text classification, we must first transform cross-lingual labeled datasets into a common semantic space, and then combine them together to obtain a larger labeled dataset. There are two options available for data transformation.

One option is machine translation, which is commonly used in cross-lingual text classification. Labeled datasets of language A can be translated into language B for text classification tasks in language B. For example, when Elfardy et al.[8] built a suite of classifiers on customer feedback, the three non-English languages in their work were translated into English using the Google Cloud Translation API. Wan et al.[9] proposed a workflow to leverage available English language resources in Chinese sentiment classification used for analyzing product reviews, their data was also translated by machine translation. Qian Sun et al.[10] built an intelligent and scalable spam detection system by learning from bilingual LinkedIn data, their large labeled dataset was also generated via machine translation. All the above studies obtained better performance leveraging cross-lingual datasets through machine translation. While the improvement in performance may due to a larger combined cross-lingual dataset through machine translation. Machine translation does introduce errors which can result in an overall lower quality training dataset. Despite this, machine translation can reduce data sparseness, as synonyms in one language most likely are unified to a common term in another language, which may also improve the performance of text classification leveraging machine translation[11].

An alternative approach is to align word embedding vector spaces trained from cross-lingual resources, and convert words into vectors in the aligned common semantic space. For example, Artetxe et al.[12], Smith et al. [13] and Joulin et al. [14] developed a series of methods to improve the quality of bilingual word vector alignment for machine translation, which can also be leveraged for cross-lingual text classification (CLTC). Goran et al. [15] proposed to exploit continuous semantic text representations and induce a joint cross-lingual semantic vector space for developing a topic classifier on a cross-lingual political document corpus, which outperforms monolingual classifiers. The core premise of this approach is to align word vectors and create a common embedding space, depending on the availability of a parallel dictionary to provide anchor points for the two original word embedding spaces. However, the variations of words in different contexts make it impossible to create an exact one-to-one alignment in a common sparse semantic space, introducing uncertainty with respect to the effect of cross-lingual data transformation with

This work was supported in part by the China Postdoctoral Science Foundation (No. 2017M622731), Natural Science Foundation of Guangdong Province (No. 2015B010106008 and Project No. 2015B010131011).

J. Jiang, S. Pang, X. Zhao and Q. Feng are with the College of Biomedical Engineering, Southern Medical University, Guangzhou, China. (email: smujiang@gmail.com, pangshumao@126.com, myjob2010@126.com, qianjinfeng08@gmail.com).

L. Wang, A. Wen and Hongfang Liu are with Department of Health Science Research, Mayo Clinic, Rochester, MN, US. (email: Wang.Liwei@mayo.edu, Wen.Andrew@mayo.edu, Liu.Hongfang@mayo.edu)

\*Q. Feng and H. Liu are corresponding authors.

the alignment approach.

In this paper, we present a comprehensive evaluation on the effects of different data transformation and combination strategies on state-of-art CNN and RNN text classification models. We first introduce our evaluation methods, then describe experiments and results followed by discussion and conclusion.

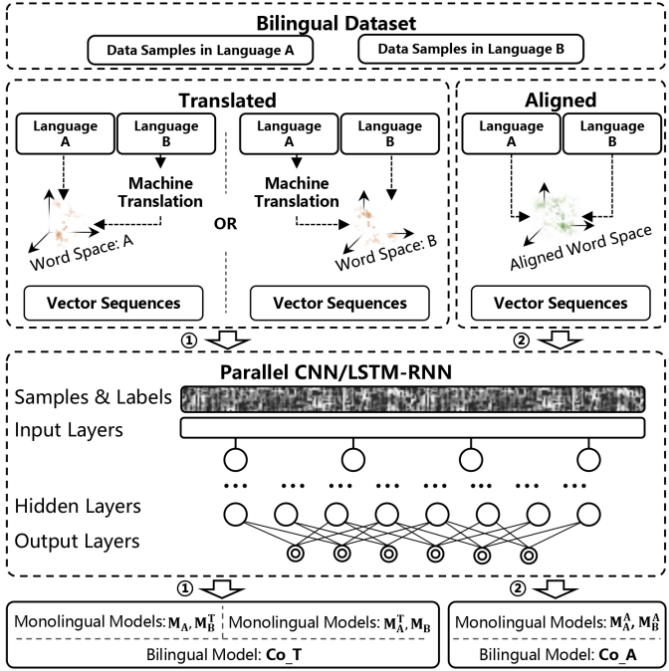

Fig. 1. Workflow of our study. Our research based on a dataset consists of language A and B. We combined our bilingual data in two different ways: 1) translate samples in language A to language B, and convert all data samples into vector sequences in a single word space for language B; 2) convert all the data samples in a same pre-aligned word space. Co_T and Co_A denote the co-trained model for two scenarios respectively. $M_A$ and $M_B$ denote the models trained from language A and language B respectively. $M_A^T$ denotes the model trained with language A' translation, while $M_B^T$ denotes the model trained with language B' translation. $M_A^A$ denotes the model trained with language A in an aligned word space, $M_B^A$ denotes the model trained with language B in the same aligned word space.

## II. METHODS

To evaluate the effect of cross-lingual data transformation and combination on monolingual/bilingual text classification models, we converted our data into vector sequences and combined them using two different methods: 1) utilizing of machine translation to unify data samples, and convert all data samples in a single word space; 2) using a pre-aligned common semantic space to represent bilingual data simultaneously. We then introduced CNN with multiple parallel convolutional channels and RNN with multiple LSTM (Long Short-Term Memory) units as our text classification models. Vector sequences obtained in two scenarios as well as corresponding labels were fed into our CNN and RNN. For each scenario, we fed A and B separately into our neural nets to train two monolingual models, and fed both A and B into the same neural nets to train a bilingual model.

### A. Word Vector Spaces

Word Embedding is an effective way to represent words into vectors for applying deep neural network approaches. Different training corpora and different embedding models will however yield different word vectors, let alone different languages in our study. In order to make our investigation to be more comprehensive and make our result to be more generalizable, we decided to choose pre-trained word vectors distributed by FastText for our study [5, 16, 17]. There are word vectors for 157 languages[1], trained on Common Crawl and Wikipedia, and these models were trained using CBOW with position-weights, in dimension 300, with character n-grams of length 5, a window size of 5 and 10 negatives. These word vectors can be used in scenarios 1 and 3.

Aligned word vectors for 44 languages are also available[2]. These word vectors can be used in scenario 2. However, vocabulary size in these words embedding is very limited. Since the alignments are performed with the RCSLS method described in Joulin et al's paper[14] [5], we adopted this method to obtain an enlarged word embedding trained from Common Crawl to meet our demands.

### B. Classification models

After word embedding, all words in our dataset were transformed into vectors, and could be regarded as the input of the classification models. Kim et al. [18] proposed a CNN model with multi-convolution channels to extract sentiment information, which has already been widely used in many text classification tasks[1, 19, 20]. In this paper, we introduced this classic CNN model for our text classifier evaluation. As a comparative study, we also introduce a RNN model into our study. The RNN is modified from the classic CNN, and it has already been used by other investigators[21]. Both neural network architectures are shown in Fig. 2. The main difference is the substitution of the convolutional and max-pooling layers of CNN with LSTM units, which can be seen in Fig. 2(a) and (c). To highlight the difference between traditional CNN model and RNN model, in particular, we demonstrated the details of convolution and max-pooling layers in Fig. 2(b) and LSTM units in Fig. 2(d). We focused on revealing the differences of two models, and some crucial parameter settings for model training.

*1) Convolutional Neural Network*

The CNN in our study is illustrated in Fig. 2, which has been widely used in many similar tasks [22] [19]. The specific feature of this model is that it uses multiple channels of convolution filters to capture semantic information hidden in sentences. This multi-channel convolutional neural network consists of 4 convolution and max-pooling parallel layers, one flattened fully connected layer, two dense layers and one dropout layer. With a fully connected softmax layer as output, we can get the probability on each classification label.

Details of the first six layers in our CNN model are illustrated in Fig. 2(b), in which $w_i$, denotes the words in sentences, were represented as 300-dimensional word vectors, and the maximum sentence length was set to 100 words, such that the input sample is a 100×300 matrix. Random noise was appended to the rear of vector sequence if a sentence was shorter than 100 words. There are 4 parallel convolution and max-pooling channels for input data feature selection, and the size of convolutional filter is 2x300, 3x300, 4x300 and 5x300. Mathematical abstraction can be performed for each unit.

---
[1]https://fasttext.cc/docs/en/crawl-vectors.html
[2]https://fasttext.cc/docs/en/aligned-vectors.html

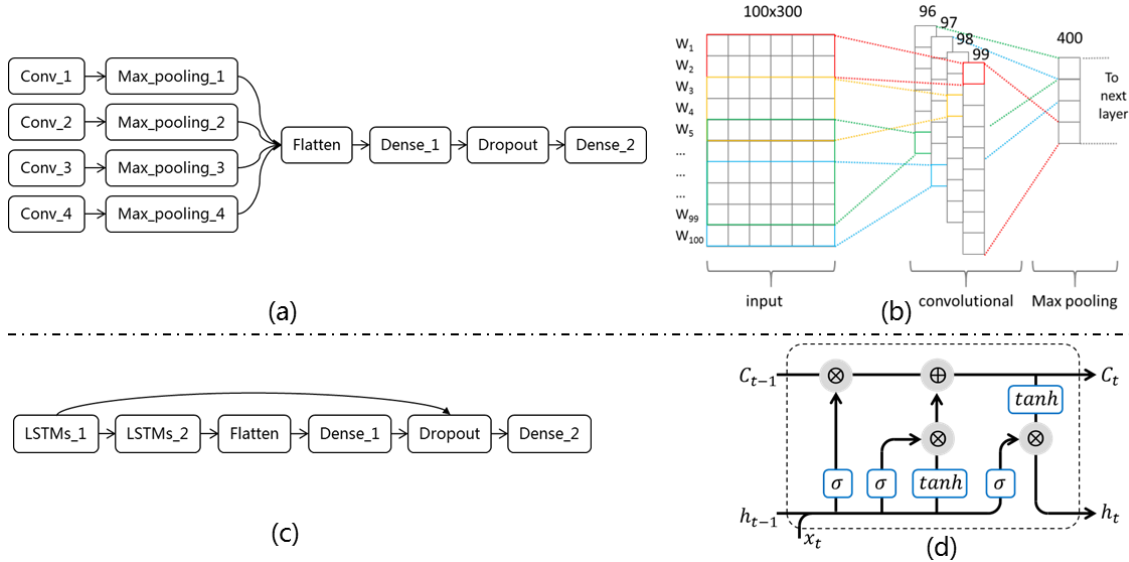

Fig. 2. Neural network structure of our text classification models. (a) Overview of CNN. (b) Details of the convolutional and max pooling layers in our CNN model. (c) Overview of RNN. (d) Details of LSTM unit in our RNN model. The differences between CNN and RNN lie in the first couple of layers.

Let $X_i \in R^k$ be a $k$-dimensional vector of the $i$-th word in the corresponding sentence. A sentence of length n can be expressed as:

$$X_{1:n} = X_1 \oplus X_2 \oplus \cdots \oplus X_n \quad (1)$$

Where $\oplus$ denotes the concatenation operator. Under a typical scenario, we use $X_{i:i+j}$ to represent the concatenation of $X_i, X_{i+1}, \cdots, X_{i+j}$.

Assuming a convolutional operation contains a filter $w \in \mathcal{R}^{hk}$, we can use this filter (size:$h \times k$) to generate new features for a sentence by operating on a sliding window. For instance, a sliding window $X_{i:i+h-1}$ generate a feature $C_i$ according to the following equation:

$$C_i = f(w * x_{i:i+h-1} + b), \quad (2)$$

where $w \in \mathcal{R}$ is an offset term and $f$ is a non-linear convolutional kernel function, such as the hyperbolic tangent or Sigmoid function. This filter is applied to all possible window positions, $\{X_{1:h}, X_{2:h+1}, \cdots, X_{n-h+1:n}\}$ to generate the characteristic mapping

$$c = [c_1, c_2, \cdots, c_{n-h+1}]. \quad (3)$$

Here, we have $c \in \mathcal{R}^{n-h+1}$. Next, each word mapping is applied with a maximum pooling operation, where the specific operation is to take $\hat{c} = \max\{c\}$ as the eigenvalue of the word mapping. The goal of this operation is to extract the most important feature for each word mapping, namely, the feature of maximum value, which can also be called semantic information.

*2) Recurrent Neural Network*

We substituted the convolutional and max-pooling layers of CNN with Long Short-Term Memory (LSTM) layers to build our RNN model. Fig. 2(c) shows the RNN structure for text classification process used in our model. Similar network architectures were also seen in [4, 21, 23].

As shown in Fig. 2(c), the network architecture is almost identical to a classic CNN model, with the exception of the first two LSTMs layers. Considering the input sample is a 100X300 matrix, there should be 100 LSTM units for each LSTM layer. The Long Short-Term Memory (LSTM) unit is very widely used in deep learning for NLP (Natural Language Processing) today to combat the vanishing gradient problem through a gating mechanism. The main advantage of LSTM is that they can learn long-term dependencies, so we usually concatenate several LSTM units to build an available structure.

The structure of a LSTM unit has been shown in Fig. 2(d), in which, $C_{t-1}$ and $h_{t-1}$ are the internal memory (cell state) and hidden state of the last LSTM unit respectively. $x_t$ is the $t$-th input vector, corresponding to the $t$-th word in a sentence for our work. The input of σ and $tanh$ function is the concatenation of $h_{t-1}$ and $x_t$. $f_t, i_t$ and $o_t$ are the forget, input and output gates, respectively. $\tilde{C}_t$ is a "candidate" hidden state that is computed based on the current input and the previous hidden state. Current cell state $C_t$ and hidden state $h_t$ can therefore be formulated as:

$$C_t = \sigma(f_t * C_{t-1} + i_t * \tilde{C}_t) \quad (4)$$
$$h_t = \tanh(C_t) * o_t \quad (5)$$

If we denote $W$ as the recurrent connection at the previous hidden layer and the current hidden layer, $U$ as the weight matrix connecting the inputs to the current hidden layer, so that we can denote:

$$f_t = \sigma(x_t U^f + h_{t-1} W^f) \quad (6)$$
$$i_t = \sigma(x_t U^i + h_{t-1} W^i) \quad (7)$$
$$o_t = \sigma(x_t U^o + h_{t-1} W^o) \quad (8)$$
$$\tilde{C}_t = tanh(x_t U^g + h_{t-1} W^g) \quad (9)$$

III. EXPERIMENTS

*A. Data Resource*

We evaluated the performance of monolingual models trained from the transformation of cross-lingual data on both AG's News[4] and Webis-CLS-10[24]. The evaluation of bilingual models trained from the combination of English and French was conducted on Webis-CLS-10.

AG's News is a dataset which contains more than 1 million news articles gathered from more than 2000 news sources. We chose the 4 largest classes (including World, Sports, Bussiness, and Technices) from this corpus to construct our dataset, using only the title and description fields. Webis-CLS-10 is a Cross-Lingual Sentiment (CLS) dataset consisting of approximately 800,000 Amazon product reviews in the four languages English, German, French, and Japanese, which is provided by Prettenhofer et al. [24] and extended by Blitzer et al.[25]. In this cross-lingual dataset, each review contains a category label, a rating, a title and a review text. We chose only the English and French reviews for our study, because they are the two most commonly used languages. There are three category labels including books, DVD and music, and data samples are evenly distributed. With the comments as input, text classifier will be trained to identify which category does the reviewer comments on.

For both datasets, the raw text data was split by category labels and divided with portion of 70%, 15% and 15% for training, validation and testing respectively. Then the raw data was shuffled and merged into training, validation and testing file. Raw text in these three files was chopped to 100 words if the reviews are long enough. Random noise was appended to the end of the sentence if the reviews are shorter than 100 words.

### B. Environment and Settings

Our models were trained on a remote workstation with two NVIDIA Titan X GPU with 12 GB VRNM, two Intel Xeon CPU E5-2623 v3 @ 3.00GHz and 8*16GB DDR4 RAM @2133MHz. Keras, a high-level neural networks API, was introduced to implement CNN and RNN network in our study. Tensorflow 1.12.1 was installed as the backend of Keras. Network architectures of both two models in our study were validated with Tensorboard.

To make our models comparable and repeatable, model parameters were fixed in the training phase. For both RNN and CNN, we used Adam as our optimizer, and dropout rate was set to 50%, batch size was set to 32, and leaning rate was set to $10^{-3}$. Early stopping patience was set to 5 and maximum training epoch was set to 300. Both English and French data were split into training, validation and testing set, with proportion of 0.70:0.15:0.15. Fully independent hold-out validation strategy was applied in the training phase to optimize our models. Categorical cross entropy was introduced as our loss function. We used mean average precision to estimate model training convergence. To avoid overfitting, both the training and validation accuracy were also monitored in Tensorboard. Multiclass weighted-average precision, recall and F1 score were calculated to quantitatively measure the performance of our model. All code and shell scripts used in this study are publicly available on GitHub[3].

## IV. RESULTS

### A. Evaluation on Monolingual Text Classification

In this section, we evaluated the performance of monolingual text classification models trained from original and transformed data. Data transformations used included both machine translation and embedding alignment.

*1) Translated*

In this experiment, we translated French and English data into each other, and compared the performance of model trained from original data and its translation. We adopted Baidu Fanyi, which is very similar with Google Translation, as the machine translator. The original data and its translation were converted into vector sequences by retrieving the corresponding word vectors.

[3]https://github.com/smujiang/Crosslingual_classification.

With consistent training settings, monolingual models were trained for both French and English. Both data set (Webis-CLS-10 and AG's News) were trained and evaluated. The evaluation results are listed in Table I, which shows French classification model can be constantly improved by translating the data into English. On contrary, when we translated English into French, and train the model in the same way, we found that the performance of the English classification model declined. This phenomenon can be observed on both RNN and CNN models.

Considering English is a more popular language than French, we can infer that monolingual models can be optimized by translating the training data into another language with richer language resources.

TABLE I
MONOLINGUAL MODEL F1 SCORE: DATA TRANSLATION

|       | RNN    |        | CNN    |        |
|-------|--------|--------|--------|--------|
|       | CLS    | AG     | CLS    | AG     |
| ENG   | 0.7809 | 0.9220 | 0.6364 | 0.7283 |
| T-ENG | 0.6980 | 0.9008 | 0.5238 | 0.6128 |
| FRE   | 0.6972 | -      | 0.5720 | -      |
| T-FRE | 0.8798 | -      | 0.6238 | -      |

*ENG denotes the model is trained and tested with original English, T-ENG denotes the model is trained and tested with translated English (French). Similarly, FRE denotes the model is trained and tested with original French, and T-FRE denotes model trained and tested with French data translation (English).

*2) Aligned*

In this experiment, we converted both English and French training data into vector sequences by querying a common representation space. RCSLS method[14] was introduced to align the pre-trained English and French word vectors and create a common semantic space. With the same training and

TABLE II
MONOLINGUAL MODEL F1 SCORE: EMBEDDING ALIGNMENT

|       | RNN    |        | CNN    |        |
|-------|--------|--------|--------|--------|
|       | CLS    | AG     | CLS    | AG     |
| ENG   | 0.7809 | 0.9220 | 0.6364 | 0.7283 |
| A-ENG | 0.7919 | 0.9253 | 0.7407 | 0.9009 |
| FRE   | 0.6972 | -      | 0.572  | -      |
| A-FRE | 0.6906 | -      | 0.6639 | -      |

*ENG denotes training and testing samples were represented in the original English embedding space, A-ENG denotes the training and testing samples were represented in the aligned common space. FRE and A-FRE are similar.

evaluation settings, monolingual models were trained and evaluated with each of the two datasets. The evaluation results are shown in Table II.

As we can observe in Table II, monolingual model trained from English can be boosted by choosing a shared common semantic space with French. However, there is a slight decrease on the A-FRE, which can be regarded as a fluctuation of model training. Specifically, performance promotion is more significant on CNN models.

### B. Evaluation on Bilingual Text Classification

In this section, we evaluated the performance of bilingual models trained from the combination of both English and French. Two data combination strategies were utilized, 1) converting words into vectors in their own representation spaces and then stacking bilingual samples together; 2) translating one language to another, converting words into vectors in the representation space of the target language, then stacking bilingual samples together.

Since this experiment requires a bilingual dataset, only Webis-CLS-10 was included in this portion. Performance of previously trained monolingual models were included in our comparison, so that we can know if there is an improvement on bilingual models trained with the combination of two languages.

As we observed that data transformation has diverse effect on the performance of monolingual model trained on English and French, we evaluated the effect of data combination of each language separately.

*1) English*

In this experiment, bilingual models trained with both English and French data based on different combination methods were compared with the monolingual model trained with only original English. All models were tested with English represented in the corresponding semantic spaces of the scenarios.

TABLE III
BILINGUAL MODEL F1 SCORE: ENGLISH

|  | RNN | CNN |
| --- | --- | --- |
| ENG | 0.7809 | 0.6364 |
| Aligned (ENG+FRE) | 0.7990 | 0.7875 |
| Translated (ENG+FRE) | 0.7696 | 0.6308 |

*ENG denotes training and testing samples were represented in the original English embedding space. All the rest are bilingual models, which were co-trained with English and French, and tested with English. Aligned (ENG+FRE): English and French were represented in a common semantic space. Translated (ENG+FRE): English were translated into French, and then represented in French embedding space.

We can observe that bilingual models may not necessarily perform better than monolingual models. Translating English into French may introduce more errors into the system, and result in a downside of the model. Otherwise, the co-trained model can benefit from bilingual dataset, and get higher evaluation metric, which is more obvious in CNN model.

Specifically, the bilingual model trained from English and French represented in the aligned space performs best, which accords with our previous evaluation of monolingual models (Table II, row "A-ENG" and Table III, row "Aligned (ENG+FRE)"). Moreover, bilingual model can reach even higher score than monolingual, which means the co-trained model can benefit from the expansion of training set.

*2) French*

The experiment design in this section is quite similar to the previous one. Bilingual models trained with both English and French data based on different combination methods were compared with monolingual model trained with only original French. All models were tested with French represented in the corresponding semantic spaces of the scenarios.

TABLE IV
BILINGUAL MODEL F1 SCORE: FRENCH

|  | RNN | CNN |
| --- | --- | --- |
| FRE | 0.6972 | 0.5720 |
| Aligned (ENG+FRE) | 0.7017 | 0.6618 |
| Translated (ENG+FRE) | 0.8880 | 0.6637 |

*FRE denotes training and testing samples were represented in the original French embedding space. All the rest are bilingual models, which were co-trained with English and French, and tested with French. Aligned (ENG+FRE): English and French were represented in a common semantic space. Translated (ENG+FRE): French were translated into English, and then represented in English embedding space.

As we can see in Table IV, bilingual models constantly perform better than the monolingual model trained with only French. As in our previous evaluation on monolingual models, machine translation can significantly promote the performance of French classification model. In this experiment, co-trained models can also benefit from the expansion of training set because we can observe a score increase on bilingual model compared to monolingual (Table I, row "T-FRE" and Table IV, row "Translated (ENG+FRE)").

## V. DISCUSSIONS

In this paper, we evaluated the performance of RNN and CNN text classifiers with differing cross-lingual data transformation and combination strategies. Our results suggest that semantic space transformation, including machine translation and embedding alignment, can be used to booster the performance of text classifiers. Specifically, translation of data into another language with richer language resources can improve performance. For the resource rich language itself, aligning the embedding space with another language can also promote the performance.

Intuitively, machine translation may introduce errors into the dataset, because some words and phrases cannot be directly translated into corresponding words or phrases, but have to be contextualized via adjacent sentences. On the other hand, machine translation may also be able to simplify language variations by creating semantic links across languages. The key lies in if the benefits can overcome the shortcomings. Our experiment results proved that modern machine translation can transform text data from a sparse semantic space into a dense space, thus text classifier can be boosted. Similarly, Farrús et al.[11] also pointed out that machine translators tend to use certain words to express various similar sentiments. Otani et al. [26] also leveraged machine translation to simplify the ambiguity of context in the specific knowledge domain to acquire knowledge projection.

Word embedding space alignment-based methods may significantly rely on the presence of a parallel dictionary to provide anchor points to train an available common word representation space. However, synonyms are very common in languages and the meaning of words change over context. Intuitively, this may introduce errors to the downstream process. Our experiments however proved that this method can also provide an effective way to improve the classification models. The benefits may be derived from the fact that word embedding can be reassigned during word alignment, links between words sharing the similar semantic context may be strengthened, and some rare words can also get more meaningful representations through word alignment. Some previous investigations declared similar views. For example, Sabet et al.[27] proposed a way of word alignment for rare words with word embedding.

Our evaluation results on bilingual models suggested that neural network-based text classification models are able to concurrently learn from bilingual datasets, and provide higher accuracy than monolingual models. The benefits may come from two aspects, one is the data transformation (machine translation and embedding alignment) step, and the other is the data sample expansion. Additionally, we observed that text classification models co-trained from cross-lingual data can get better performance even if the network is simple. Some previous investigators also claimed that the performance of their text classification models was boosted by co-training on bilingual dataset. For example, Kaiser et al. [28] developed a complicate attention-based network to tackle multiple tasks classification, and they found tasks with less data benefit largely from joint training with other tasks, while performance on large tasks degrades only slightly if at all. However, in this paper, we found that artificial neural network with simple architecture designed for text classification task can also benefit from leveraging cross-lingual data.

One limitation of our research is that we only evaluated the CNN and RNN models on very limited cross-lingual dataset. However, the evaluation metrics are obvious enough to support our conclusion. More comprehensive evaluation can be conducted if more public datasets for cross-lingual text classification are available. Furthermore, with our pipeline, more languages can be included into the assessment, rather than just English and French.

## VI. CONCLUSIONS

In this paper, we evaluated the performance of two commonly used CNN and RNN text classifiers with different data transformation and combination strategies for text classification tasks.

Our evaluation results suggested that monolingual models trained from one language can benefit from another language by translating or aligning it into another semantic space if the target space has richer language resources. We also observed that text classification models can significantly benefit from co-training on cross-lingual datasets by converting them into a same semantic space with both machine translation and embedding alignment.

Meanwhile, we observed that co-trained models can achieve better performance even if the architecture of deep neural network is simple.